\documentclass[12pt]{iopart}
\usepackage[dvips]{graphicx,color}
\begin{document}

\def\mmbf#1{\mbox{\boldmath${#1}$}}
\def\smD{{\scriptscriptstyle D }}
\def\smC{{\scriptscriptstyle {\rm C} }}
\def\smQ{{\scriptscriptstyle {\rm Q} }}
\def\smB{{\scriptscriptstyle {\rm B} }}
\def\smLP{{\scriptscriptstyle {\rm LP} }}
\def\smLB{{\scriptscriptstyle {\rm LB} }}

\title[$dE/dx$ to Subleading Order]{$\mmbf{dE/dx}$ to Subleading 
Order in the Plasma Density}

\author{Lowell S. Brown, Dean L. Preston, and Robert L. 
Singleton Jr.}

\address{Los Alamos National Laboratory, Los Alamos, 
New Mexico 87545, USA}

\eads{\mailto{brownl@lanl.gov}, \mailto{bobs1@lanl.gov}, 
\mailto{dean@lanl.gov}}

\begin{abstract}
  Dimensional
  continuation is employed to compute the energy loss
  rate for a non-relativistic particle moving through
  a highly ionized plasma. No restriction is made on
  the charge, mass, or speed of this particle, but it
  is assumed that the plasma is not strongly coupled
  in that the dimensionless plasma coupling parameter 
  $g= e^2 \kappa_\smD / 4 \pi T $ is small, where 
  $\kappa_\smD$ is the Debye wave number.  To leading 
  order in this coupling, $dE/dx$ is of the generic form 
  $g^2 \, \ln[ g^2 C] $.  The prefactor of the logarithm 
  is well known. We compute the constant $C$ under the 
  logarithm exactly. Our result differs from approximations 
  given in the literature, with differences in the range 
  of about 20\% for cases relevant to inertial confinement 
  fusion experiments.
\end{abstract}

\vskip-0.4cm 
\pacs{52.25.Fi} 

\vskip-0.56cm \vbox{\hfill  LA-UR-05-6527}

\section{$\mmbf{dE/dx}$ and the Coulomb Log}

The stopping power of plasma component $b$ for projectile $p$ 
is of the form
\begin{eqnarray}
\label{spitzerform}
  \frac{dE_b}{dx} 
  &=& 
  \frac{e_p^2 \, e_b^2}{4\pi}\, 
  \frac{n_b}{m_b\,v_p^2}\, \ln \Lambda_b 
  =
  \frac{e_p^2}{4\pi} \, 
  \frac{\kappa_b^2}{\beta_b m_b v_p^2}\ln\Lambda_b \ ,
\end{eqnarray}
where the Coulomb logarithm $\ln\Lambda_b$ involves a ratio 
of short- and long-distance length scales. To compute 
$\ln\Lambda_b$, we employ the method of \hbox{\em dimensional 
continuation} \cite{BPS}.  
To introduce this method, we consider the Coulomb potential 
$\phi_\nu(r)$ of a point source in $\nu$ spatial dimensions:
$\phi_\nu(r) \sim 1/r^{\nu-2}$. 
Clearly the long and short distance behavior of $\phi_\nu$ 
depends on the spatial dimensionality $\nu$. In high $\nu$, 
short distance (hard) interactions are accentuated, while 
in low $\nu$ the large distance (soft) physics predominates. 

For Coulomb interactions, $\nu = 3 $ is special 
in that neither hard nor soft processes are dominant. For 
$\nu < 3$, the soft physics is predominant, and for $\nu > 3$ 
the hard processes are dominant. The energy loss for $\nu
> 3$, $d E^\smB / d x$, is obtained from the Boltzmann (B)
equation, and it contains a pole $\, (\nu-3)^{-1} \,$ that reflects 
an infrared divergence in the scattering process when $\, \nu 
\rightarrow 3^+ \,$. The energy loss for $\nu<3$, $d E^\smLB/ 
d x$, is obtained from the Lenard-Balescu (LB) kinetic equation,
and it contains a pole $\, (3-\nu)^{-1} \,$ that reflects an 
ultraviolet divergence when $\, \nu \rightarrow 3^- \,$. 
The stopping power to subleading order is therefore  
\begin{eqnarray}
  \frac {d E}{d x} = \lim_{\nu \rightarrow 3} 
  \left(\frac {d E^\smLB} {d x} + \frac {d E^\smB}{d x} 
  \right) \, , 
\label{e1}
\end{eqnarray}
and it is completely finite. Hence the two poles must cancel.
The dependence of the residues of the poles on $\, \nu \,$ 
brings in a logarithm of the ratio of the relevant short- 
and long-distance length scales, which is precisely
the Coulomb logarithm. 

\section{Collective Excitations: Lenard-Balescu 
Equation for $\mmbf{\nu < 3}$}

The soft physics is described to leading order in the plasma 
density by 
\begin{equation}
  {\partial \over \partial t} f_a({\bf p}_a) = -
  {\sum}_b \, { \partial \over \partial {\bf p}_a }
  \cdot {\bf J}_{ab} \ ,
\label{lbeq}
\end{equation}
which is the Lenard--Balescu kinetic equation for plasma species 
$a$ and $b$, where
\begin{eqnarray}
\fl
  {\bf J}_{\!ab} \!=\! e_{\!a}^2 e_{\!b}^2 \!\!\int \!\!{ d^\nu{\bf k}
  \over (2\pi)^{\nu} } { {\bf k} \over ({\bf k}^2)^2  }
  { \pi \over \left| \epsilon ({\bf k} , \!{\bf v}_{\!a} \!\cdot\!
  {\bf k} ) \right|^2 }
  \!\int \!\! { d^\nu\!{\bf p}_{\!b} \over (2\pi\hbar)^{\nu} } \,
  \delta\Big({\bf k} \!\cdot\! ({\bf v}_{\!a} \!-\! 
  {\bf v}_{\!b}) \Big)\, {\bf k} \!\cdot\!\! \left[ { \partial \over
  \partial {\bf p}_b} - { \partial \over
  \partial {\bf p}_{\!a}}  \right] \!f_{\!a}({\bf p}_{\!a}) 
  f_{\!b}({\bf p}_{\!b})   ,
\label{lbj}
\end{eqnarray}
with $ {\bf v}_a = {\bf p}_a / m_a$ and ${\bf v}_b = {\bf p}_b / m_b $.  
The collective behavior of the plasma enters through its dielectric 
function 
\begin{equation}
  \epsilon({\bf k},\omega) = 1 + {\sum}_c \, {e_c^2
  \over k^2 } \int { d^\nu {\bf p}_c \over
  ( 2\pi\hbar )^\nu } { 1 \over \omega - {\bf k} \cdot 
  {\bf v}_c + i \eta}\, {\bf k} \cdot {\partial
  \over \partial {\bf p}_c } f_c({\bf p}_c)\, , \;\; \eta \to 0^+ .
\label{epsilon}
\end{equation}

The rate of kinetic energy loss of species $a$ to species $b$ 
is given by 
\begin{eqnarray}
  {d {\cal E}^\smLB_{ab} \over dt} &=& 
\int { d^\nu{\bf p}_a \over
  (2\pi\hbar)^\nu}\, {p_a^2 \over 2 m_a } { \partial \over 
  \partial {\bf p}_a } \cdot {\bf J}_{ab} \,.
\label{dEabdt}
\end{eqnarray}
We evaluate this for the case in which species $a$ is a single 
projectile of mass $m_p$ and velocity ${\bf v}_p$, 
$
  f_a({\bf p}_a) = (2\pi\hbar)^\nu \, \delta^{(\nu)}
  ( {\bf p}_a - m_p {\bf v}_p ) \, ,
$
and the distribution function $f_b({\bf p}_b)$ for plasma species 
$b$ is Maxwell-Boltzmann at temperature $ T_b = 1 / \beta_b $. 
With $dx = v_p dt$,
\begin{eqnarray}
\fl
\nonumber
\frac{dE^\smLB_{b}}{dx} 
 =
 \frac{e_p^2}{4\pi}\,  \frac{1}{\beta_b m_p v_p^2}\, 
 \frac{\Omega_{\nu -2}}{2\pi}\, \left(\frac{K}{2\pi}\right)^{\nu-3}
 \hskip-0.3cm \frac{1}{3-\nu} \int_{0}^1 du \, (1-u)^{(\nu-3)/2} \,
  \rho_b(v_p u^{1/2}\,) 
\Bigg[\beta_b M_{pb}\, v_p^2 
  - \frac{1}{u} \Bigg] 
\\[5pt] \nonumber 
\fl
  \hskip1.2cm + {e_p^2 \over 4 \pi } { i \over 2 \pi }
  \int_{-1}^{+1} \hskip-0.25cm d\!\cos\theta \, \cos\theta \, 
  {\rho_b (v_p\cos\theta) 
  \over \sum_c \rho_c (v_p\cos\theta) } F(v_p \cos\theta) 
  \ln \hskip-0.1cm \left( { F(v_p \cos\theta) \over K^2 }\right)
\\[5pt]
\fl
  \hskip1.2cm
  -{e_p^2 \over 4 \pi } { i \over 2 \pi } { 1 \over 
  \beta_b m_p v_p^2 } {\rho_b(v_p) \over \sum_c \rho_c (v_p) }
  \Bigg[ F(v_p ) \ln  \hskip-0.1cm \left( { F(v_p ) \over K^2 } 
  \right) - F^*(v_p)\ln  \hskip-0.1cm \left( { F^*(v_p) \over K^2 } 
  \right) \Bigg] \, ,
\label{lesssingglessregg}
\end{eqnarray}
where $M_{pb}=m_p+m_b$ is the total mass, 
$F(v)=k^2\left[\,\epsilon({\bf k},k v)-1\, \right]$, and
$\rho_b(v)=\kappa_b^2 \sqrt{\beta_b m_b/2\pi}\, v 
\exp(-\beta_b m_b v^2/2)$. Here $\Omega_\nu$ is the area 
of a unit sphere in $\nu$ dimensions and $K$ is an
arbitrary wave number whose dependence cancels in
the limit (\ref{e1}).

\section{Hard Collisions: Boltzmann Equation for 
$\mmbf{\nu > 3}$}

Hard collisions in the plasma density are correctly described 
by the Boltzmann equation, which gives
\begin{eqnarray}
 { d E^\smB_b \over dx} &=& {1\over v_p} \, \int {d^\nu{\bf p}_b 
 \over (2\pi\hbar)^\nu} f_b({\bf p}_b) \, v_{pb} \, \int d 
 \sigma_{pb} \,{1\over2} m_p \left[ {v_p'}^2 - v_p^2 \right] \,,
\label{eloss}
\end{eqnarray}
where $ v_{p b} \!=\! \vert {\bf v}_p \,\!-\!\, {\bf v}_b 
\vert$, and $d\sigma_{pb}$ is
the full quantum-mechanical differential cross section for
scattering of the projectile ($p$) from the initial velocity 
${\bf v}_p = {\bf p}_p / m_p$ 
to the final velocity ${\bf v}_p'$ off a plasma particle ($b$).  
Straightforward kinematical manipulations exploiting the axial
symmetry of the scattering produce the form
\begin{eqnarray}
 { d E^\smB_b \over dx} &=& {1 \over v_p} \, \int {d^\nu{\bf p}_b 
 \over (2\pi\hbar)^\nu} f_b({\bf p}_b) \, { {\bf P} \cdot {\bf p}
 \over 2 p^2 \, M_{pb} } \, v_{pb} \, \int  d \sigma_{pb} \, 
 \, q^2  \,,
\label{eeloss}
\end{eqnarray}
in which ${\bf P}$ is the total momentum of the center of mass, 
${\bf p}$ is the relative momentum in the center of mass, and 
${\bf q}$ is the momentum
transfer.

The {\em classical} cross section in $ \nu $ dimensions
is $d \sigma_{pb}^\smC = \Omega_{\nu -2} \,  B^{\nu -2} dB$,
where $B$  is the classical impact parameter. Some calculation 
gives
\begin{equation}
 \int \, d \sigma^\smC_{pb} \, q^2 = \frac{\Omega_{\nu - 2}}{2\,\pi} 
 \frac {m^2_{pb}} {p^2} \frac {(e_p e_b)^2} {2 \, \pi} \left[ 
 \frac {p^{2 (\nu - 3)}} {\nu - 3} - \ln \left(\frac {e_p \, 
 e_b \, m_{pb}} {4} \right) - \gamma \right] \, ,
\label{dsigma-qsqrd}
\end{equation}
with $m_{pb}=m_p m_b/M_{pb}$ being the reduced mass. Placing 
the result~(\ref{dsigma-qsqrd}) in Eq.~(\ref{eeloss}) yields
\begin{eqnarray}
 && \frac{dE^\smB_{b \,\smC}}{dx} 
  = 
  \frac{e_p^2}{4\pi}\,\frac{1}
  {\beta_b m_p v_p^2} \int_0^1 \hskip-0.1cm du \,   
  \rho_b(v_p\sqrt{u}\,) \, 
  \left\{ \left[ \frac{\Omega_{\nu-2}}{2\pi}\,\frac{1}{\nu-3}
  (1-u)^{(\nu-3)/2} \right. \right. 
\nonumber\\ && 
  + \left. \left. 2 - 2\gamma - \ln\hskip-0.1cm \left(
  \frac{e_p e_b \beta_b m_b}{2m_{pb}}\,\frac{u}{1-u} 
  \right) \right] 
  \Bigg( \beta_b M_{pb}\, v_p^2 - 
 \frac{1}{u} \Bigg ) + \frac{2}{u} \, \right\} \ .
\label{288bobsnotes}
\end{eqnarray}

Making the decomposition $\int d \sigma_{pb}\, q^2=
\int d \sigma^\smC_{pb}\, q^2 + \int ( d \sigma_{pb} - 
d \sigma^\smC_{pb} ) \, q^2$ expresses $dE_b^\smB/dx =
dE^\smB_{b\,\smC}/dx + dE^\smB_{b\, \smQ}/dx$, where
$dE^\smB_{b\, \smQ}/dx$ is the {\em quantum} correction
to Eq.~(\ref{288bobsnotes}). The integral $ \int ( d \sigma_{pb} - 
d \sigma^\smC_{pb} ) \, q^2 $ is most easily evaluated by first 
calculating $ \int ( d \sigma_{pb} - d \sigma^\smB_{pb} ) \, 
q^2 $, where $ d \sigma^\smB_{pb} $ is the Born approximation 
to $ d \sigma_{pb} $, and then subtracting the contribution
$ \int ( d \sigma^\smC_{pb} - d \sigma^\smB_{pb} ) \, 
q^2 $. Inserting the correction $ v_{pb} \, 
\int ( d \sigma_{pb} - d \sigma^\smC_{pb} ) \, q^2 $ into 
Eq.~(\ref{eeloss}) yields
\begin{eqnarray}
\nonumber
  \frac{d E^\smB_{b \,\smQ}}{dt} &=& \frac{e_p^2\,\kappa_b^2}{4\pi}
  \frac{2 v_p}{\sqrt{2\pi \alpha_b}}~ e^{-\alpha_b/2} 
  \!\!\int_0^\infty \! du\, \Bigg\{e^{-\alpha_b u^2/2} \Bigg[ 
  \ln(\eta_b/u) - {\rm Re}\,\psi(1 + i \eta_b/u) \,\Bigg]
\\[5pt] && \qquad 
  \left[\,\frac{M_{pb}}{m_p\,u} \left(\cosh\alpha_b u - 
  \frac{\sinh \alpha_b u}{\alpha_b u}\right) - 
  \frac{m_b}{m_p} \sinh\alpha_b u \, \right] \Bigg\} \ ,
\label{dedtqmcalf}
\end{eqnarray}
where $\psi$ is the logarithmic derivative of the gamma
function, ${\rm Re}$ denotes the real part, $\alpha_b \equiv 
\beta_b m_b v^2_p$\,, and $\eta_b \equiv {e_p e_b} / {4\pi\hbar 
v_p}$.

\section{Results}

\begin{figure}
\centerline{
\includegraphics[height=8cm]{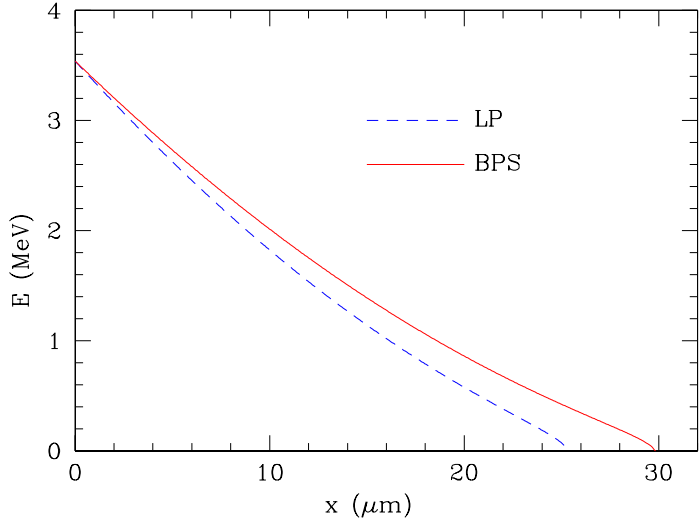}
}
\vskip-1cm 
\caption{
The energy $E(x)$ (in MeV) of an $\alpha$ particle with 
initial energy $E_0=3.54\,{\rm MeV}$ vs. the distance $x$ 
(in $\mu\,{\rm m}$) traveled through an equal molal $DT$ 
plasma. Note that $E(x)$ is obtained by inverting $x=
\int_{E_0}^E dE \, (dE/dx)^{-1}$, where the stopping 
power $dE/dx$ has been expressed as a function of energy. 
The plasma temperature is $T=3\, {\rm keV}$ and the electron 
number 
density is $n_e=10^{25}\, {\rm cm}^{-3}$. The plasma coupling 
is small, $g=0.011$, and so our calculation (BPS) is essentially 
exact. Our result is shown by the solid curve. The work of Li 
and Petrasso \cite{LIP} is often used in laser fusion simulations.
Their result (LP) is shown by the dashed line. Note that the difference 
in the total ranges between our result and that of Li and 
Petrasso of about $5 \,\, \mu m$ is a little larger than 
20\%.
}
\label{fig:aDT030125Exfig}
\end{figure}

\begin{figure}
\centerline{
\includegraphics[height=8cm]{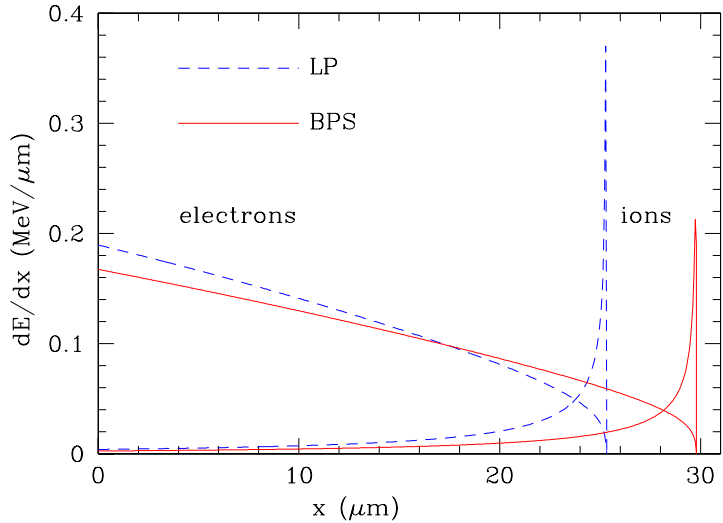}
}
\vskip-1cm 
\caption{
The $\alpha$ particle $dE(x)/dx$ (in ${\rm Mev}
/\mu{\rm m}$) vs. $x$ (in $\mu m$) split into 
separate ion (spiked curves) and electron components 
(softly decreasing curves). The area under each curve 
gives the corresponding energy partition into electrons 
and ions. For our results (BPS), the total energy 
deposited into electrons is $E_e=3.16$~MeV and into ions 
is $E_I=0.38$~MeV, while LP give $E_e^\smLP=3.11$~MeV 
and $E_I^\smLP=0.43$~MeV. These energies sum 
to the initial $\alpha$ particle energy of $E_0=3.54\,
{\rm MeV}$. Note that BPS has a longer $\alpha$ particle 
range and deposits less energy into ions than LP. Both
observations would tend to make fusion more difficult 
to achieve for BPS than for LP. 
}
\label{fig:aDT030125dEdxxeifig}
\end{figure}

The total stopping power is the sum of the contributions from 
large-distance collective excitations $ dE^\smLB / dx $ and 
from short-distance hard collisions $ dE^\smB / dx $, that is, 
the sum over species $b$ of Eqs.~(\ref{lesssingglessregg}), 
(\ref{288bobsnotes}), and (\ref{dedtqmcalf}). The poles at 
$\nu = 3$  and the $\ln K$ terms cancel. Our result for 
$dE/dx$ is generically of the form $n\,(\ln n + C)$ in the
plasma density $n$, and it is accurate to all orders in the
quantum parameter $\eta_b$. Figures~\ref{fig:aDT030125Exfig}
and \ref{fig:aDT030125dEdxxeifig}  illustrate our result with 
an example that is relevant to the DT plasmas in laser fusion 
capsules. 

\end{document}